%% file: icissp23.tex
\documentclass[a4paper,twoside]{article}

\usepackage{epsfig}
\usepackage{subcaption}
\usepackage{calc}
\usepackage{amssymb}
\usepackage{amstext}
\usepackage{amsmath}
\usepackage{amsthm}
\usepackage{multicol}
\usepackage{pslatex}
\usepackage{apalike}
\usepackage[bottom]{footmisc}
\usepackage{SCITEPRESS}     

\usepackage{graphicx}
\usepackage{xspace}
\usepackage{hyperref}
\usepackage{esvect}
\usepackage{wrapfig}
\usepackage{url}
\usepackage{booktabs}
\usepackage[inline]{enumitem}
\usepackage{listings}
\usepackage{prftree}
\usepackage{proof}
\usepackage{bussproofs}
\usepackage{xcolor}
\usepackage{textcomp}
\usepackage{tikz}
\usepackage{microtype}
\usepackage{tikz}
\usepackage{float}
\usepackage{multirow}
\usepackage{booktabs}
\usepackage{arydshln}
\usepackage{caption}
\usepackage{subcaption}
\usepackage[none]{hyphenat}

\begin{document}
\sloppy

\title{Technical Report: Automating Vehicle SOA Threat Analysis using a Model-Based Methodology}


\author{\authorname{Yuri Gil Dantas\sup{1}, Simon Barner\sup{1}, Pei Ke\sup{2}, Vivek Nigam\sup{2} and Ulrich Sch\"{o}pp\sup{1}}
\affiliation{\sup{1}fortiss GmbH, Munich, Germany}
\affiliation{\sup{2}Huawei Technologies D\"{u}sseldorf GmbH, D\"{u}sseldorf, Germany}
\email{\{lastname\}@fortiss.org, \{first.lastname\}@huawei.com}
}


\input{macros}

\abstract{
While the adoption of Service-Oriented Architectures (SOA) eases the implementation of features such as autonomous driving and over-the-air updates, it also increases the vehicle's exposure to attacks that may place road-users in harm. 
To address this problem, standards (ISO 21434/UNECE) expect manufacturers to produce security arguments and evidence by carrying out appropriate threat analysis. 
As key threat analysis steps, e.g., damage/threat scenario and attack path enumeration, are often carried out manually and not rigorously, security arguments lack precise guarantees, e.g., traceability w.r.t. safety goals, especially under system updates. 
This article proposes automated methods for threat analysis using a model-based engineering methodology that provides precise guarantees with respect to safety goals. 
This is accomplished by proposing an intruder model for automotive SOA which together with the system architecture and the loss scenarios identified by safety analysis are used as input for computing assets, impact rating, damage/threat scenarios, and attack paths. To validate the proposed methodology, we developed a faithful model of the autonomous driving functions of the Apollo framework, a widely used open source autonomous driving stack. 
The proposed machinery automatically enumerates several attack paths on Apollo, including attack paths not reported in the literature.
} 

\onecolumn \maketitle \normalsize \setcounter{footnote}{0} \vfill

\noindent\paragraph*{Remark:} A shorter version of this technical report has been accepted for publication at ICISSP 2023.

\input{sec-introduction}

\input{sec-apollo_modeling}

\input{sec-intruder_model}

\input{sec-automation}
\input{sec-related_work}


\input{sec-conclusion}
\bibliographystyle{apalike}
{\small
\bibliography{references}}



\end{document}

%% file: macros.tex
\makeatletter
\newcommand\CO[1]{%
  \@tempdima=\linewidth%
  \advance\@tempdima by -2\fboxsep%
  \advance\@tempdima by -2\fboxrule%
  \leavevmode\par\noindent%
  \fbox{\parbox{\the\@tempdima}{%
    \small\normalfont\sffamily \textcolor{red}{#1} }}%
  \smallskip\par}
\makeatother

\newcommand{\red}[1]{\textcolor{red}{#1} }

\newcommand{\eg}{{e.g.}}
\newcommand{\ie}{{i.e.}}
\newcommand{\etal}{\emph{et al.}}
\newcommand{\cf}{{cf.}}

\newenvironment{scprooftree}[1]%
  {\gdef\scalefactor{#1}\begin{center}\proofSkipAmount \leavevmode}%
  {\scalebox{\scalefactor}{\DisplayProof}\proofSkipAmount \end{center} }

\newcommand\tmr{\texttt{TMR}\xspace}
\newcommand\hmd{\texttt{HmD}\xspace}
\newcommand\htd{\texttt{HtD}\xspace}
\newcommand\safmon{\texttt{MonAct}\xspace}
\newcommand\sa{\texttt{SA}\xspace}
\newcommand\watchdog{\texttt{WD}\xspace}
\newcommand\sanity{\texttt{SanChk}\xspace}
\newcommand\nprog{\texttt{NProg}\xspace}

\newcommand\soamachinery{\textsc{LAUFEN}\xspace}

\newcommand\machineryOld{\textsc{SafPat}\xspace}
\newcommand\machinery{\textsc{SafSecPat}\xspace}

\newcommand\tool{\textsc{Pattern Synthesis}\xspace}

\newcommand\before{\ensuremath{\mathsf{before}}}
\newcommand\dsl{\ensuremath{\mathsf{SafPat}}}
\newcommand\subcp{\ensuremath{\mathsf{subcp}}}
\newcommand\ch{\ensuremath{\mathsf{ch}}}
\newcommand\info{\ensuremath{\mathsf{if}}}
\newcommand\hw{\ensuremath{\mathsf{hw}}}
\newcommand\sw{\ensuremath{\mathsf{sw}}}
\newcommand\dep{\ensuremath{\mathsf{dep}}}
\newcommand\id{\ensuremath{\mathsf{id}}}
\newcommand\interface{\ensuremath{\mathsf{interface}}}
\newcommand\ecu{\ensuremath{\mathsf{ecu}}}
\newcommand\Dep{\ensuremath{\mathsf{dep}}}
\newcommand\hz{\ensuremath{\mathsf{hz}}}
\newcommand\fl{\ensuremath{\mathsf{fl}}}
\newcommand\ft{\ensuremath{\mathsf{ft}}}
\newcommand\dm{\ensuremath{\mathsf{dm}}}
\newcommand\flToHz{\ensuremath{\mathsf{fl2Hz}}}
\newcommand\ftTofl{\ensuremath{\mathsf{ft2fl}}}
\newcommand\tp{\ensuremath{\mathsf{fl_{tp}}}}
\newcommand\sev{\ensuremath{\mathsf{pt_{sv}}}}
\newcommand\trsev{\ensuremath{\mathsf{th_{sv}}}}
\newcommand\subhz{\ensuremath{\mathsf{subHz}}}
\newcommand\err{\ensuremath{\mathsf{err}}}
\newcommand\loss{\ensuremath{\mathsf{loss}}}
\newcommand\omission{\ensuremath{\mathsf{omission}}}
\newcommand\late{\ensuremath{\mathsf{late}}}
\newcommand\early{\ensuremath{\mathsf{early}}}
\newcommand\minor{\ensuremath{\mathsf{min}}}
\newcommand\major{\ensuremath{\mathsf{maj}}}
\newcommand\hazardeous{\ensuremath{\mathsf{haz}}}
\newcommand\cat{\ensuremath{\mathsf{cat}}}
\newcommand\safMon{\ensuremath{\mathsf{safMon}}}
\newcommand\sm{\ensuremath{\mathsf{sm}}}
\newcommand\watchDog{\ensuremath{\mathsf{watchDog}}}
\newcommand\wdcp{\ensuremath{\mathsf{wd}}}
\newcommand\nProg{\ensuremath{\mathsf{nProg}}\xspace}
\newcommand\dProg{\ensuremath{\mathsf{2Prog}}\xspace}
\newcommand\voter{\ensuremath{\mathsf{Voter}}}
\newcommand\VOTER{\ensuremath{\mathsf{VT}}}
\newcommand\fs{\ensuremath{\mathsf{fs}}}
\newcommand\candsl{\ensuremath{\mathsf{can}}}

\newcommand\sg{\ensuremath{\mathsf{sg}}}
\newcommand\fop{\ensuremath{\mathsf{f_{op}}}}
\newcommand\fsilent{\ensuremath{\mathsf{f_{sl}}}}
\newcommand\fsafe{\ensuremath{\mathsf{f_{sf}}}}

\newcommand\allNFail{\ensuremath{\mathsf{allXfail}}}
\newcommand\mostNFail{\ensuremath{\mathsf{mostXfail}}}
\newcommand\never{\ensuremath{\mathsf{never}}}

\newcommand\idd{\ensuremath{\mathsf{\id_{d}}}}
\newcommand\idfl{\ensuremath{\mathsf{\id_{fl}}}}
\newcommand\idft{\ensuremath{\mathsf{\id_{ft}}}}
\newcommand\idhz{\ensuremath{\mathsf{\id_{hz}}}}
\newcommand\system{\ensuremath{\mathsf{sys}}}
\newcommand\hzsev{\ensuremath{\mathsf{hz_{sev}}}}
\newcommand\hzexp{\ensuremath{\mathsf{hz_{exp}}}}
\newcommand\hzctl{\ensuremath{\mathsf{hz_{ctl}}}}
\newcommand\idc{\ensuremath{\mathsf{\id_{cp}}}}
\newcommand\idhw{\ensuremath{\mathsf{\id_{hw}}}}
\newcommand\idcOne{\ensuremath{\mathsf{\id_{c1}}}}
\newcommand\idcTwo{\ensuremath{\mathsf{\id_{c2}}}}
\newcommand\idhwOne{\ensuremath{\mathsf{\id_{hw1}}}}
\newcommand\idhwTwo{\ensuremath{\mathsf{\id_{hw2}}}}

\newcommand\securityPattern{\ensuremath{\mathsf{securityPattern}}}
\newcommand\safetyPattern{\ensuremath{\mathsf{safetyPattern}}}
\newcommand\safetyPatternID{\ensuremath{\mathsf{\id}}}
\newcommand\safetyPatternNAME{\ensuremath{\mathsf{name}}}
\newcommand\safetyPatternCP{\ensuremath{\mathsf{cp}}}
\newcommand\safetyPatternINP{\ensuremath{\mathsf{inp}}}
\newcommand\safetyPatternINT{\ensuremath{\mathsf{int}}}
\newcommand\safetyPatternOUT{\ensuremath{\mathsf{out}}}
\newcommand\safetyIntent{\ensuremath{\mathsf{safetyIntent}}}
\newcommand\securityIntent{\ensuremath{\mathsf{securityIntent}}}

\newcommand\idpt{\ensuremath{\mathsf{\id_{pt}}}}
\newcommand\idt{\ensuremath{\mathsf{\id_{th}}}}
\newcommand\idsg{\ensuremath{\mathsf{\id_{sg}}}}
\newcommand\mcs{\ensuremath{\mathsf{mcs}}}
\newcommand\idmcs{\ensuremath{\mathsf{\id_{msc}}}}
\newcommand\lmcsToHz{\ensuremath{\mathsf{lmcs2hz}}}
\newcommand\idlmcs{\ensuremath{\mathsf{\id_{lmcs}}}}

\newcommand\asil{\ensuremath{\mathsf{asil}}}
\newcommand\asila{\ensuremath{\mathsf{a}}}
\newcommand\asilb{\ensuremath{\mathsf{b}}}
\newcommand\asilc{\ensuremath{\mathsf{c}}}
\newcommand\asild{\ensuremath{\mathsf{d}}}

\newcommand\publicdsl{\ensuremath{\mathsf{public}}}
\newcommand\potThreat{\ensuremath{\mathsf{pThreat}}}
\newcommand\threatdsl{\ensuremath{\mathsf{threat}}}
\newcommand\reachI{\ensuremath{\mathsf{reachI}}}
\newcommand\pathI{\ensuremath{\mathsf{P}}}
\newcommand\bdCh{\ensuremath{\mathsf{bdCh}}}
\newcommand\ttpdsl{\ensuremath{\mathsf{pt_{tp}}}}
\newcommand\trtype{\ensuremath{\mathsf{th_{tp}}}}
\newcommand\severe{\ensuremath{\mathsf{sev}}}
\newcommand\moderate{\ensuremath{\mathsf{mod}}}
\newcommand\negligible{\ensuremath{\mathsf{neg}}}
\newcommand\authenticity{\ensuremath{\mathsf{auth}}}
\newcommand\availability{\ensuremath{\mathsf{avl}}}
\newcommand\integrity{\ensuremath{\mathsf{int}}}
\newcommand\confidentiality{\ensuremath{\mathsf{con}}}

\newcommand\hls{\texttt{HLS}\xspace}
\newcommand\cam{\texttt{C\scriptsize{AM}}\xspace}
\newcommand\cell{\texttt{C\scriptsize{ELL}}\xspace}
\newcommand\bt{\texttt{B\scriptsize{T}}\xspace}
\newcommand\navig{\texttt{N\scriptsize{AVIG}}\xspace}
\newcommand\gw{\texttt{G\scriptsize{W}}\xspace}
\newcommand\obd{\texttt{OBDC\scriptsize{onn}}\xspace}
\newcommand\bdctl{\texttt{B\scriptsize{D}\normalsize{C}\scriptsize{TL}}\xspace}
\newcommand\hlswt{\texttt{H\scriptsize{L}\normalsize{S}\scriptsize{WT}}\xspace}
\newcommand\pwrswtact{\texttt{P\scriptsize{WR}\normalsize{S}\scriptsize{WT}}\xspace}
\newcommand\canbus{\texttt{CAN Bus}\xspace}
\newcommand\can{\texttt{CAN}\xspace}
\newcommand\canI{\texttt{CAN Bus 1}\xspace}
\newcommand\canII{\texttt{CAN Bus 2}\xspace}
\newcommand\canIII{\texttt{CAN Bus 3}\xspace}

\definecolor{mGreen}{rgb}{0,0.6,0}
\definecolor{mGray}{rgb}{0.5,0.5,0.5}
\definecolor{mPurple}{rgb}{0.58,0,0.82}
\definecolor{backgroundColour}{rgb}{0.95,0.95,0.92}

\lstdefinestyle{CStyle}{
    backgroundcolor=\color{backgroundColour},   
    commentstyle=\color{mGreen},
    keywordstyle=\color{magenta},
    numberstyle=\tiny\color{mGray},
    stringstyle=\color{mPurple},
    basicstyle=\scriptsize,
    breakatwhitespace=false,         
    breaklines=true,                 
    captionpos=b,                    
    keepspaces=true,                 
    numbers=left,                    
    numbersep=5pt,                  
    showspaces=false,                
    showstringspaces=false,
    showtabs=false,                  
    tabsize=2,
    language=C
}

%% file: sec-introduction.tex
\section{Introduction}

The automotive industry is under great transformation to meet challenges of implementing features such as Autonomous Driving and Over-the-Air Updates. 
  Instead of using distributed architectures with domain-specific hardware, vehicles are using software-intensive \emph{Service-Oriented Architectures} (SOA) with powerful centralized computer units.
  The open-source Apollo framework~\cite{apollo} is an example of this transformation providing autonomous vehicle features that have been used in the development of real-world autonomous vehicle applications, such as autonomous taxis and buses.  
  
This transformation has also increased concerns on how attackers can affect road-user safety. While security threats to safety have been known for more than a decade ago~\cite{attack.jeep}, the upcoming/recent standards ISO 21434~\cite{iso21434} and the UNECE~\cite{unece} have pushed industry to change its development process to enable safe and secure-by-design vehicles. For example, the ISO 21434 puts great emphasis on the development process and on the threat analysis, \eg, Damage/Threat Scenario/Attack Path enumeration, that shall be performed and addressed before putting the vehicle on the road.
At the end, Original Equipment Manufactures (OEMs) shall provide compelling arguments and evidence, \ie, an assurance case, that their vehicles are safe also from a security perspective.

OEMs may pay a costly price if they develop autonomous vehicle features without previously producing analysis, argument, and evidence supporting vehicle safety and security. 
Without these artifacts, it is hard to expect that these vehicles will be accepted by certification agencies and be allowed to be used in several countries, once standards are more heavily enforced. Even more troublesome is that several attacks have been reported that can cause serious hazards to road-users, such as vehicle collisions. 
\emph{As we claim here, many of these attacks could have been identified during the design of the system architecture by using a safe and secure-by-design approach with suitable threat analysis supported by automation.} 

A key challenge for the development of safe and secure-by-design vehicles is handling the enormous complexity involved. 
For example, without adequate countermeasures, SOA allows any software component to publish any data including data that may be consumed by safety-critical functions. 
This has been a source of, \eg, overprivilege attacks~\cite{HongKJCCMM20} causing hazardous situation whenever a safety-critical function consumes data erroneously published by a malicious component (or even by a faulty component).
For another example, malicious components may exploit SOA communication vulnerabilities to cause man-in-the-middle attacks~\cite{ZelleLKK21}.
Moreover, sensors, such as cameras and GPS radios, are attack surfaces that may be exploited by attackers to cause hazards~\cite{JhaCBCTKI20,Shen20}. 


\subsection{Safe and Secure-by-Design Methodology and Contributions}
\label{subsec:method}
\begin{figure*}[t]
\begin{center}
\includegraphics[width=\textwidth]{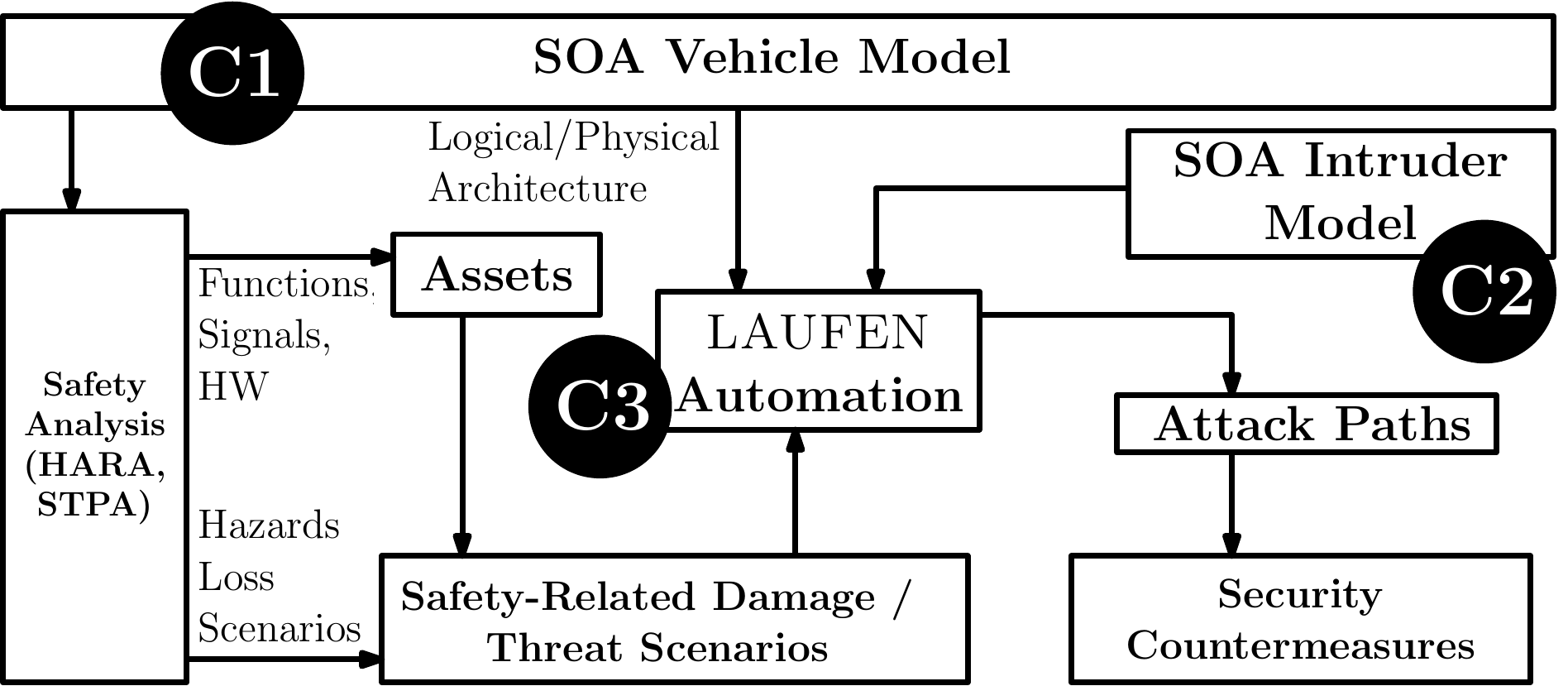}
\end{center}
\caption{Illustration of the proposed safe and secure-by-design methodology, tool-chain and key contributions (C1, C2 and C3).}
\label{fig:approach}
\end{figure*}


The proposed safety and security methodology and three key contributions are depicted in Figure~\ref{fig:approach}.
The methodology is built upon the following key ideas from the automotive safety and security co-engineering literature:
\begin{itemize}
  \item Analysis Techniques for Software-Intensive Systems: System Theoretic Process Analysis (STPA)~\cite{stpa-handbook} has been recommended for safety analysis of autonomous functions by standards such as the ISO 21448 (Safety Of The Intended Functionality -- SOTIF)~\cite{sotif} for assuring the safety of features such as autonomous driving.
  This is because STPA does not assume linear causal dependency and rather puts a greater emphasis on the faulty/malicious component interactions. 
  \item Safety to Security: The approach recommended by Bosch engineers~\cite{bosch-paper} uses safety artifacts, \eg, safety goals and hazards, as inputs to security analysis. 
  There are two key motivations for this: 
  1) A safety analysis is typically carried out before a security analysis. 
  2) By using safety as input to security, one can claim, through appropriate traceability, completeness of security analysis w.r.t. to the results of the safety analysis. This is done, for example, by checking whether all causes of hazards (called loss scenarios in STPA terminology) have traces to appropriate security analysis.
  \item Model-Based Tool-Chains: Model-based engineering approaches are based on formal abstractions of the system under design and therefore help mitigate the complexity of nowadays software and hardware architectures and to boost development speed and quality when compared to traditional document-based approaches by means of automated analysis, design and validation tools.
\end{itemize}

While these methods have been proposed, this article is the first to apply them together into an overarching model-based methodology for SOA vehicle architectures. 
As depicted in Figure~\ref{fig:approach}, we start from a (SOA) Vehicle Model, specifying the key functions, logical components, and platform (a.k.a. physical) architecture. 
These model elements ensure the soundness of the approach, as the safety and security analysis that follow are traced to the model.
From the Hazard Analysis and Risk Asessment (HARA) and STPA analysis, key safety functions, channels and physical elements are identified, which are then traced as assets from the security perspective that need to be protected.
Loss scenarios obtained from STPA, \ie, the situations that may lead to hazards, are traced to damage and threat scenarios specifying how intruders can cause safety hazards.
From this point onward, we carry out a security analysis, \eg, using the logical and platform architectures to identify attack paths that can cause threat scenarios.
Ultimately, we discuss potential countermeasures to address threats.

The key benefits of the approach are three-fold: The first benefit is a full traceability between safety and security analysis and the vehicle model. 
This means that the analysis is reflected in the actual implementation that will be deployed in the vehicle. The second benefit is that the methodology provides guarantees that all loss scenarios for all hazards are considered by the security analysis, \eg, all loss scenarios are traced to damage/threat scenarios.   
This means that all identified safety issues shall be considered from the security perspective. 
The third benefit is that our model-based methodology enables the use of automated methods, \eg, the automated enumeration of attack paths based on intruder models. 

The main contributions of this article are:
\begin{itemize}
  \item \textbf{Apollo-Based Vehicle Model (C1):} By examining the relevant pieces of code in the Apollo code-base related to autonomous driving functions, we designed a faithful vehicle model. 
   The model reflects the SOA publish and subscribe pattern, and the information (namely the topics) between the Apollo components.
  To the best of our knowledge, it is the first model based on the Apollo v7.0.0 code base.
  
  \item \textbf{Intruder Model for Vehicle SOA (C2):} By examining vehicle SOA security literature, we formalized an intruder model for vehicle SOA.
    The intruder is capable of carrying out Man-in-the-Middle (MITM) attacks, and carrying out spoofing attacks by infiltrating the system from public interfaces to, \eg, exploit perception sensors, such as LiDAR and Camera. 

  \item \textbf{Attack Path Automation (C3):} 
  We developed a machinery (\soamachinery) to automate the enumeration of attack paths on the vehicle system architecture.
  \soamachinery takes as input the model, assets, damage/threat scenarios, and the implementation of the intruder model, and outputs all attack paths.
\end{itemize}

We demonstrate and validate our approach and automation on the developed Apollo Vehicle Model.
Our focus is on safety assets as it is the main concern for autonomous driving.
The developed machinery identified \textbf{246} attack paths. The attack paths include attacks that have been reported in the literature.
Given the traceability to safety analysis, our machinery identifies a much greater number of attack paths that would need to be mitigated (or for which some security rational shall be provided) by security countermeasures.
Indeed, based on the generated attack paths, we identified potential attacks that have not yet been reported.


%% file: sec-apollo_modeling.tex
\begin{figure*}[t]
\centering
\includegraphics[width=\textwidth]{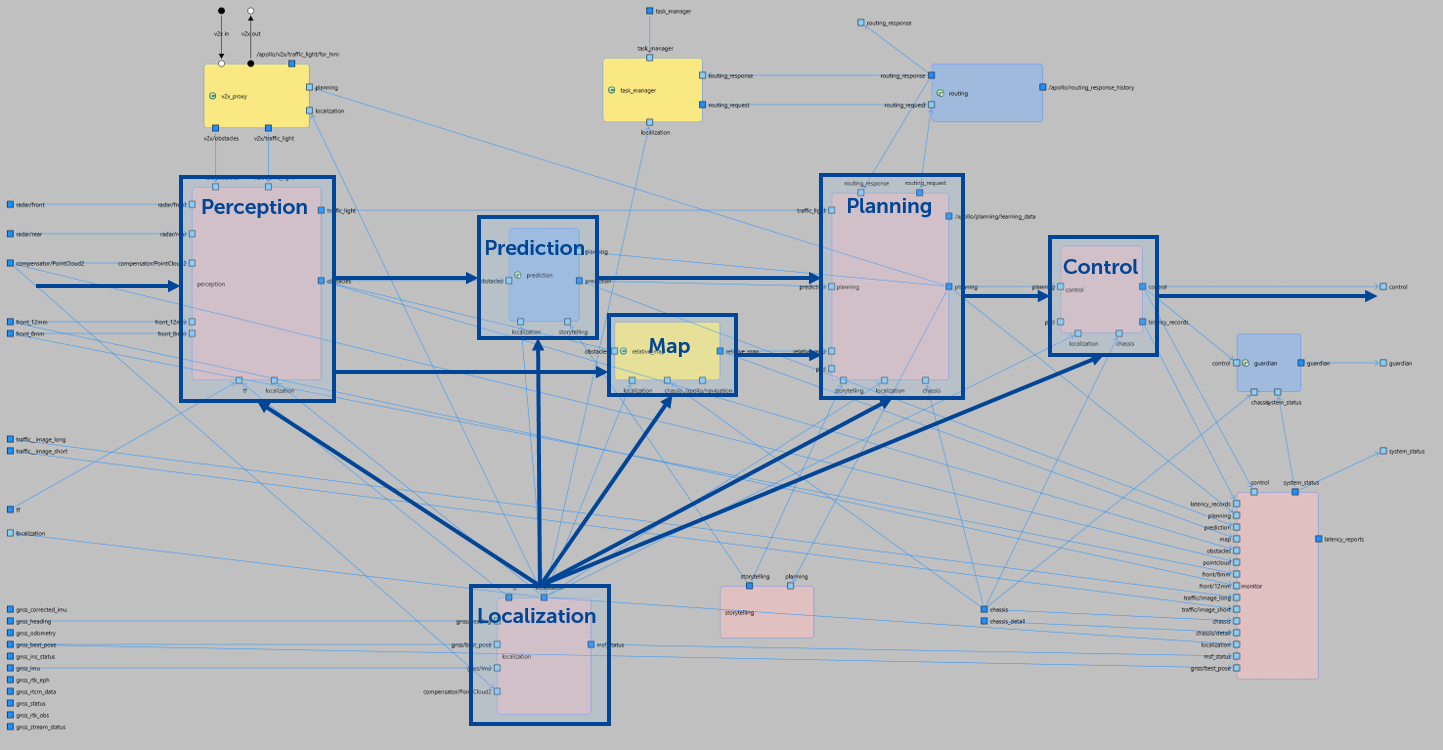}
\caption{Logical architecture: Main autonomous driving components}
\label{fig:la_apollo_main_system}
\end{figure*}

\section{Apollo Modeling}\label{sec:apollo-modeling}

Apollo~\cite{apollo} is an open-source autonomous driving stack enabling highly autonomous vehicle features 
(more precisely, at Level 4 in the SAE ranking~\cite{sae}), such as Highway and Traffic Jam Pilots, 
where a vehicle can drive with limited human supervision.
Apollo v7.0.0~\cite{apollo} consists of more than 500k lines of \texttt{C++} code.

A central part of the Apollo implementation is the Cyber RT middleware~\cite{cyberrt}.
Cyber RT provides a publish/subscribe pattern to enable the communication between software components running over it.
Components can communicate via tagged channels, a.k.a. topics.
Components may publish data to topics by writing messages to a named topic and may subscribe to any topic of interest by referring to the topic name.
Whenever a publisher writes data to a topic, this data is received by all subscribers. 
Cyber RT allows more than one component to publish data on a topic, and more than one component to subscribe to it. 
The announcement of (new) topics and the subscription of components to topic names are performed by a mechanism called service discovery.

This section describes the designed Apollo model used to demonstrate our methodology.
The model focuses on the parts of the code-base that are related to autonomous vehicle features, namely, sensors (Camera, LiDAR),
localization, perception, prediction, planning, control, and HMI. 

The Apollo system architecture has been modeled in the model-based system engineering tool AutoFOCUS3~\cite{af3}.
The model comprises of \textbf{9} functions, \textbf{61} logical components, \textbf{341} ports, transmitting \textbf{73} data structures with \textbf{361} members, \textbf{16} execution units, \textbf{12} transmission units, and \textbf{6} sensors. 
We developed an experimental metamodel~\cite{Aravantinos2015} extension in AutoFOCUS3 to describe publish/subscribe communication by means of dedicated \textit{topic} port data types.
Due to the lack of space, the remainder of this section describes only selected parts of the logical and platform architecture, since the security results presented in this article mainly focus on the logical architecture and platform architecture.

\paragraph{Logical Architecture.}
The designed logical architecture is complex and consists of four hierarchical levels with multiple components.
Figure~\ref{fig:la_apollo_main_system} depicts the second highest level of our Apollo model containing the main autonomous driving components.


    The \textbf{localization} component receives sensor data from GNSS and computes the vehicle's position. The vehicle's position is received by the following components. 
    The \textbf{perception} component receives sensor data from cameras, radars and LiDAR, and the vehicle's position. Perception identifies obstacles, such as other vehicles on the road, as well as the state of traffic lights.
    The \textbf{prediction} component takes the list of obstacles from perception and the vehicle's position, and tries to predict the intention of obstacles, which may be other vehicles or pedestrians. The prediction includes aspects such as whether a vehicle intends to change lanes.
    The \textbf{relative map} component aggregates the list of obstacles and combines it with map data, which contains information about the road, such as lanes and traffic lights.
    The \textbf{planning} component takes as input all the data computed by localization, perception, prediction and relative map. Planning uses this data to plan a safe and comfortable trajectory for the vehicle.
     The \textbf{control} component receives the planned trajectory and produces control commands (steering, acceleration, etc.) for the vehicle to follow the trajectory.

\begin{figure*}[t]
\centering
\includegraphics[width=\textwidth]{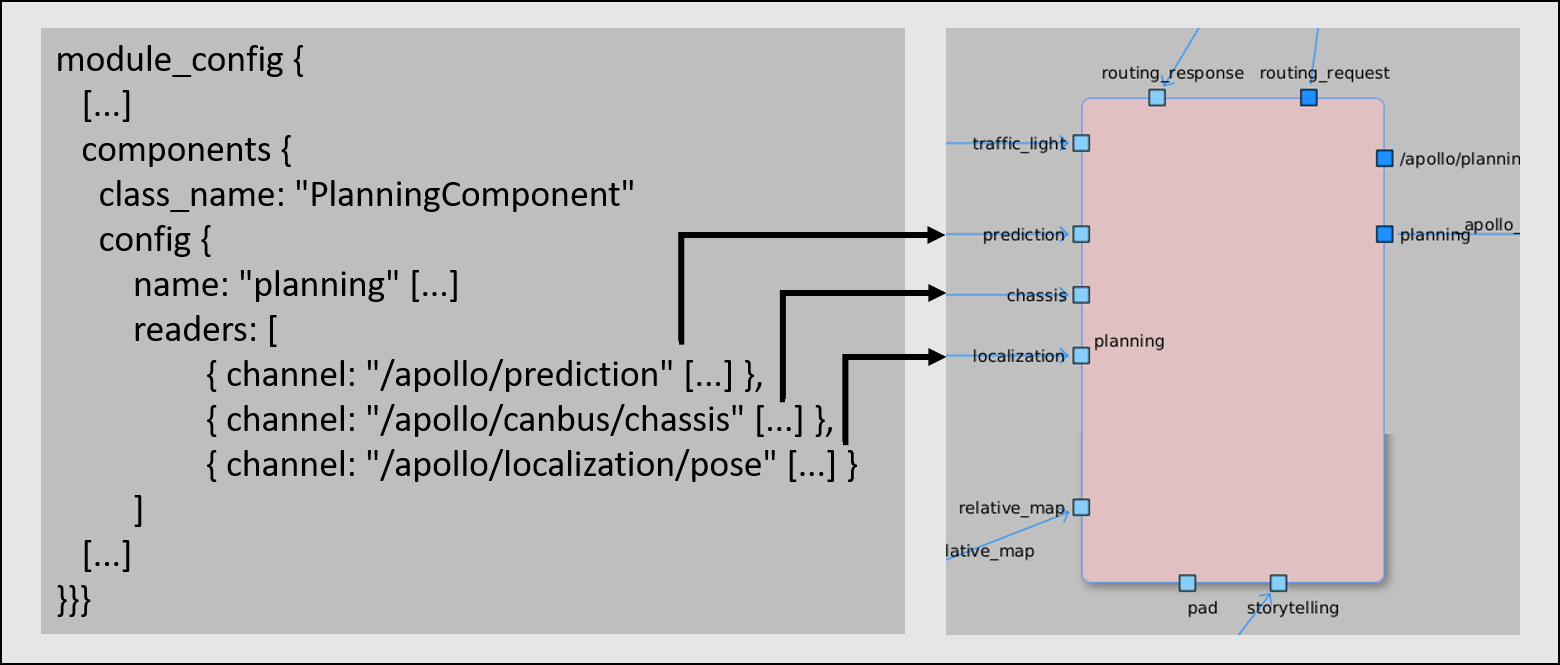}
\caption{Modeling planning's subscriber ports from its DAG configuration file.}
\label{fig:DAG_planning_code}
\end{figure*}

A key challenge was to ensure the faithfulness of the model to the Apollo code. 
To accomplish this, we extracted the model elements by manually inspecting the Apollo code.
For example, to find all Cyber RT components implemented in Apollo, we inspect the code to find all implementations of the \texttt{class cyber::Component}. 
The next step was to identify the topics and which components publish to them and subscribe to them.
The Apollo implementation specifies the topic communication using the following mechanisms: DAG configuration files, C++ code implementing readers for topics and producers of topics, and library code. 
We inspected each of these mechanisms to map the topics that are subscribed and published to components.
For example, Figure~\ref{fig:DAG_planning_code} illustrates the DAG configuration file for the planning component.
It shows that planning subscribes to the topics "\texttt{/apollo/prediction}", "\texttt{/apollo/canbus/chassis}", and "\texttt{/apollo/localization/pose}".



\begin{figure*}[t]
\centering
\includegraphics[width=\textwidth]{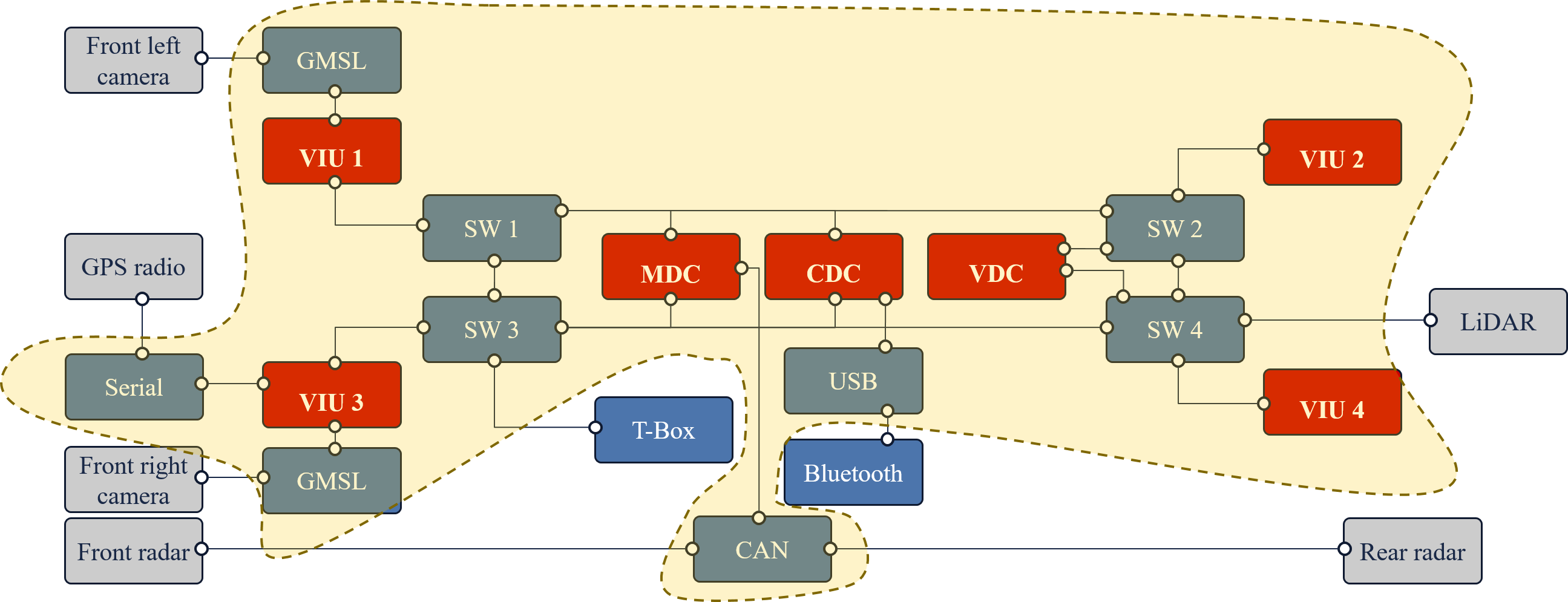}
\caption{Platform architecture based on modern smart car architecture (yellow shading represents the system boundary).}
\label{fig:pl_ecus_network}
\end{figure*}

\paragraph{Platform Architecture.}
Figure~\ref{fig:pl_ecus_network} illustrates our platform architecture that follows the trend for modern smart car architectures consisting of a few, but powerful ECUs and using network interfaces (\ie, switches) between ECUs.

The main ECUs in the platform architecture are: 
\begin{enumerate*}[label=(\arabic*)]
    \item \textbf{MDC:} Mobile Data Center: This hardware is responsible for the autonomous function related components, such as inferring objects from camera input, predicting the movement of objects in the environment, planning trajectories. 
    The MDC is further sub-divided into sub-systems with different types of processing units with different levels of safety assurance levels, such as an ASIL-D MCU.
    \item \textbf{CDC:} Intelligent Cockpit: This hardware is responsible for all the cockpit related functions, such as driver monitoring systems and entertainment functions.
    \item \textbf{VDC:} Vehicle Controller: This hardware is responsible for the basic vehicle control functions, such as Electric Power Steering, Battery Management, and Anti-lock Braking functions.
    \item \textbf{VIU 1-4:} Vehicle Integration Units: These hardware are powerful gateways that interface the MDC, CDC, and VDC, connected through network interfaces,  to the domain specific hardware connected through CAN buses.
\end{enumerate*}

The yellow shade in the model represents the system boundary (a.k.a. item boundary).
We consider as part of the system all components that are implemented in the Logical Architecture.
For example, Sensors (\eg, LiDAR and GPS radio) are not part of the system itself. 
They are third-party devices that are connected to the system and provide inputs from the environment. 
We consider them as public interfaces that are outside and may be accessed by external users.

\begin{table*}[t]
\centering
  \begin{tabular}{p{2.5cm}p{4cm}|p{5.5cm}p{0.5cm}}
    \toprule
    \textbf{Hazard HZ1} & \multicolumn{3}{p{9.5cm}}{Unintended distance between the ego vehicle and other objects.} 
    \\
    \midrule
    \textbf{Severity} & {Life-Threatening} (S3) &\\
    \textbf{Exposure} & {High Probability} (E4) & \quad \textbf{Safety Risk Level:} ASIL D\\
    \textbf{Controllability} & {Difficult to Control} (C3) &\\
  \end{tabular}
  \begin{tabular}{p{2cm}p{2cm}p{2cm}p{2.5cm}}
    \toprule
    \multicolumn{4}{p{12.8cm}}{\textbf{Short Description of Loss Scenario LS1 that causes Hazard HZ1}}
    \\  \multicolumn{4}{p{12.8cm}}{A failure occurs in planning, causing an erroneous output of planning. Planning erroneously provides the planned trajectory to control affecting the behavior of control, which may compute a target deceleration insufficient to avert a front-end collision.} \\    
    \midrule
    \textbf{Source} & \textbf{Target} & \textbf{Message} & \textbf{Failure Mode}\\
    planning & control &  trajectory  & erroneous\\
    \bottomrule
  \end{tabular} 
  \caption{Example of safety analysis results.}
  \label{table:apollo-hazard}
\end{table*}

\section{Safety-informed Security Analysis}\label{sec:safety_analysis}

Our main focus is to identify assets, damage and threat scenarios related to safety as it is the main concern to autonomous driving.
We describe how key safety artifacts are consumed by security analysts to identify key security artifacts, establishing a traceability between security and safety concerns.
One can argue from such traces the (relative) completeness with respect to safety of the security analysis in the sense that threats that can cause any one of the safety loss scenarios are identified. 
As there is existing literature that advocates similar traceability between safety and security~\cite{bosch-paper,dantas_journal_safety_security_patterns}, albeit not using loss scenarios and artifacts mentioned in the ISO 21434~\cite{iso21434}, we simply exemplify the method on examples using the Apollo system architecture. 

\paragraph{Safety Analysis.} 
We carried out a safety analysis for the Apollo system architecture.
In compliance to ISO 26262-3~\cite{iso26262}, we have identified hazards by using Hazard Analysis and Risk Assessment (HARA). Furthermore, we use System Theoretic Process Analysis (STPA)~\cite{stpa-handbook} to identify how such hazards may occur.

Relevant for this article are hazards and loss scenarios provided, respectively, by HARA and STPA.
A \emph{hazard} is a potential source of loss (\eg, loss of life) caused by malfunctioning behavior of the item (\ie, Apollo system architecture).
A \emph{loss scenario} describes the casual factors that may lead to a hazard. 

We have identified \textbf{4} hazards and \textbf{21} loss scenarios.
We will use the hazard (HZ1) and loss scenario (LS1) described in Table~\ref{table:apollo-hazard} to demonstrate the model-based methodology for threat analysis described in Section~\ref{subsec:method}. 
HZ1 is a high risk level (ASIL D) related to the autonomous driving functions.
LS1 is a possible cause for HZ1. 
LS1 is traced to two components in the model, planning and control, and to the topic containing the trajectory produced by planning.
LS1 specifies that if the computed trajectory is erroneous, \eg, instead of recommending a low acceleration, it recommends a right acceleration, HZ1 may occur, \ie, the vehicle may collide with obstacles.

\paragraph{Assets and Damage/Threat Scenarios from Safety Analysis.}
Following the ISO 21434~\cite{iso21434}, \emph{assets} are objects (\eg, software components, hardware units) for which the compromise of its cybersecurity property can lead to the damage of the item. 
A \emph{damage scenario} denotes the adverse consequence due to the compromise of a cybersecurity property of an asset.
A \emph{threat scenario} denotes the potential actions (or simply attack) on assets that can lead to damage scenarios.
The hazards and loss scenarios obtained from the safety analysis can be directly used to identify such security artifacts related to safety-related damages.

\textbf{Damage scenarios} are traced to hazards. 
The damage scenario traced to HZ1 specifies that unintended distance shall be avoided also from a security perspective.
There are three main \textbf{assets} that can be traced to the loss scenarios: \emph{Safety Functions:} The safety related functions (typically implemented as pieces of software) shall be protected. For LS1, the functions planning and control are such assets. 
\emph{Topic/Messages:} The safety-related signals/messages mentioned in the loss scenarios shall be protected. For LS1, the topic carrying the trajectory information shall be protected. \emph{Hardware/Physical:} The hardware in which safety functions are deployed shall be protected. The functions associated to LS1 are deployed at the MCU (inside of MDC) hardware unit.
Moreover, the failure mode of loss scenarios indicates which cyber-security properties (CIA properties) are associated to the these assets. 
The failure mode erroneous and loss indicate, respectively, that the integrity and availability of the corresponding assets shall be ensured.
Notice that the confidentiality property cannot be extracted from safety analysis as lack of confidentiality does not lead to safety-related damages. 
From the loss scenario and its derived assets, one can elaborate \textbf{threat scenarios} by using, \eg, the STRIDE methodology~\cite{shostack14book}. For example, the integrity of safety functions and of physical assets can be violated by tampering attacks, while the integrity of topic/messages can be violated by spoofing and elevation of privilege. 

These artifacts are used to enumerate attack paths that shall be considered, namely those that can lead to threat scenarios.
The enumeration of attack paths depends on the technology that is being used. 
For example, if a software may be updated using Over-the-Air mechanisms, then attack paths shall consider how these mechanisms can be exploited to tamper the software with malicious updates.
For the Apollo system architecture considered in this article, one needs to consider the use of SOA machinery, \eg, protocols for service discovery, publish-subscribe communication patterns, sensors, and other public interfaces, \eg, Bluetooth and WiFi. 
These are considered in the next section.

%% file: sec-intruder_model.tex
\newcommand\attackinside{\mathsf{at\_ins}}
\newcommand\attackoutside{\mathsf{at\_out}}
\newcommand\inductivecasetwopublic{\mathsf{reach\_rd}}
\newcommand\inductivecaseonepublic{\mathsf{reach\_wrt}}
\newcommand\basicpublic{\mathsf{basic\_out}}
\newcommand\basicsoa{\mathsf{basic\_ins}}
\newcommand\basicsoasub{\mathsf{basic\_ins\_sub}}
\newcommand\inductivecaseonesoa{\mathsf{reach\_ins\_rd}}
\newcommand\inductivecasetwosoa{\mathsf{reach\_2\_soa}}
\newcommand\public{\mathsf{publico}}
\newcommand\publico{\mathsf{po}}
\newcommand\vint{\vdash_I}

\newcommand\mywrite{\mathsf{write}}
\newcommand\myread{\mathsf{read}}

\newcommand\informationflow{\mathsf{if}}

\newcommand\port{\mathsf{p}}
\newcommand\portone{\mathsf{p1}}
\newcommand\porttwo{\mathsf{p2}}

\newcommand\wrt{\mathsf{wrt}}
\newcommand\rdg{\mathsf{rd}}
\newcommand\eo{\mathsf{eo}}
\newcommand\co{\mathsf{co}}
\newcommand\coone{\mathsf{co1}}
\newcommand\cotwo{\mathsf{co2}}
\newcommand\ei{\mathsf{ei}}
\newcommand\ci{\mathsf{ci}}
\newcommand\ecuo{\mathsf{ecuo}}
\newcommand\ecui{\mathsf{ecui}}
\newcommand\alloc{\mathsf{alloc}}
\newcommand\cp{\ensuremath{\mathsf{c}}}
\newcommand\cpo{\mathsf{cpo}}
\newcommand\cpone{\mathsf{c1}}
\newcommand\cptwo{\mathsf{c2}}
\newcommand\cpi{\mathsf{cpi}}
\newcommand\net{\mathsf{net}}
\newcommand\neti{\mathsf{neti}}
\newcommand\neto{\mathsf{neto}}
\newcommand\nti{\mathsf{ni}}
\newcommand\nto{\mathsf{no}}
\newcommand\reach{\mathsf{i\_reach}}
\newcommand\attack{\mathsf{i\_attack}}
\newcommand\trans{\mathsf{ch}}
\newcommand\elem{\mathsf{el}}
\newcommand\elemone{\mathsf{el1}}
\newcommand\elemtwo{\mathsf{el2}}
\newcommand\outputport{\mathsf{out}}
\newcommand\inputport{\mathsf{inp}}
\newcommand\subscriber{\mathsf{sub}}
\newcommand\topic{\mathsf{tp}}
\newcommand\topicone{\mathsf{tp1}}
\newcommand\publisher{\mathsf{pub}}
\newcommand\sensoro{\mathsf{snsro}}
\newcommand\sensor{\mathsf{snsr}}
\newcommand\so{\mathsf{so}}
\newcommand\protection{\mathsf{pro}}
\newcommand\canpublish{\mathsf{publish}}

\section{Intruder Model for Vehicle SOA}\label{sec:intruder-model}

\begin{figure*}[t]
\[
\begin{array}{l}
\textbf{Write and Read Rules}\\[2pt]

 \infer[\mywrite_1]{\Gamma \vdash \wrt(\eo,\nti)}
             {\cpo(\cp,\co),\alloc(\co,\eo), \ecuo(\ecu,\eo),\neti(\net,\nti),\trans(\eo,\nti) \in \Gamma}

\\\\[1.5pt]
 \infer[\mywrite_2]{\Gamma \vdash \wrt(\ei,\eo)}
             {\ecui(\ecu,\ei), \ecuo(\ecu,\eo), \trans(\ei,\eo) \in \Gamma}
 \quad
 \infer[\mywrite_3]{\Gamma \vdash \wrt(\nti,\nto)}
             {\neti(\net,\nti), \neto(\net,\nto), \trans(\nti,\nto) \in \Gamma}

 \\\\[1.5pt]
 \infer[\mywrite_4]{\Gamma \vdash \wrt(\publico,\nti)}
             {\public(\elem,\publico), \neti(\net,\nti), \trans(\publico,\nti) \in \Gamma}
 \quad
  \infer[\myread_1]{\Gamma \vdash \rdg(\ci,\co)}
             {\subscriber(\cpone,\ci,\topic), \publisher(\cptwo,\co,\topic) \in \Gamma}             

  \\\\[1.5pt]
  \infer[\myread_2]{\Gamma \vdash \rdg(\ei,\nto)}
             {\cpi(\cp,\ci), \alloc(\ci,\ei), \ecui(\ecu,\ei), \neto(\net,\nto),\trans(\nto,\ei) \in \Gamma}

 \\\\[1.5pt]
 \textbf{Intruder Reachability Rules}\\[2pt]
 \infer[\basicpublic]{\Gamma \vdash \reach(\publico)}
                {\public(\elem,\publico) \in \Gamma}  
\qquad \qquad \qquad \quad
 \infer[\basicsoa]{\Gamma \vdash \reach(\co)}
                {\publisher(\cp,\co, \topic) \in \Gamma}


 \\\\[1.5pt] 
 \infer[\inductivecaseonepublic]{ \Gamma \vdash \reach(\porttwo) }
                {\Gamma \vdash \wrt(\portone,\porttwo)  & \Gamma \vdash \reach(\portone)  }
\quad 
 \infer[\inductivecasetwopublic]{\Gamma \vdash \reach(\porttwo)}
                {\Gamma \vdash \rdg(\porttwo,\portone) & \Gamma \vdash \reach(\portone)   }                                
 \\\\[1.5pt]
 \infer[\inductivecaseonesoa]{\Gamma \vdash \reach(\ci)}
                {\publisher(\cp,\co,\topic), \subscriber(\cp,\ci,\topicone), \publisher(\cpone,\coone,\topicone) \in \Gamma & \Gamma \vdash \rdg(\ci,\coone) & \Gamma \vdash \reach(\co)   }






 \\\\[1.5pt]
 \textbf{Intruder Attack Rules}\\[2pt]
 \infer[\attackoutside]{ \Gamma \vdash \attack(\topic) }
                {\informationflow(\ecu,\port,\topic) \in \Gamma  & \Gamma \vdash \reach(\port)  }

  \\[3pt]
 \infer[\attackinside]{\Gamma \vdash \attack(\topic)}
                {\subscriber(\cpone,\ci,\topic), \publisher(\cp,\co,\topic), \neg \protection(\topic) \in \Gamma & \Gamma \vdash \reach(\ci) & \Gamma \vdash \reach(\co)}
\end{array}    
\]
\caption{Intruder model for SOA.}
\label{fig:intruder_capabilities}
\end{figure*}

\begin{table}[t]
  \begin{tabular}{p{1.7cm}p{5.3cm}}
    \toprule
    \textbf{Predicate} & \textbf{Denotation}\\
    \midrule
    $\ecui(\ecu,\ei)$ & ECU $\ecu$ and its input port $\ei$.\\
     \midrule
    $\ecuo(\ecu,\eo)$ & ECU $\ecu$ and its output port $\eo$.\\
     \midrule
    $\neti(\net,\nti)$ & net. interface $\net$ and its input port $\nti$.\\
     \midrule
    $\neto(\net,\nto)$ & net. interface $\net$ and its input port $\nto$.\\
     \midrule
    $\trans(\outputport,\inputport)$ & channel from output port $\outputport$ to input port $\inputport$.\\
     \midrule    
    $\wrt(\elemone,\elemtwo)$ & element $\elemone$ writes data to $\elemtwo$.\\
     \midrule
    $\rdg(\elemone,\elemtwo)$ & element $\elemone$ reads data from $\elemtwo$.\\    
     \midrule
    $\cpi(\cp,\ci)$ & component $\cp$ and its input port $\ci$.\\
     \midrule
    $\cpo(\cp,\co)$ &  component $\cp$ and its output port $\co$.\\
     \midrule
    $\alloc(\elem,\ecu)$ &  element $\elem$ is allocated to $\ecu$.\\
     \midrule
    $\publisher(\cp,\co,\topic)$ & component $\cp$ publishers the topic $\topic$ through output port $\co$.\\
     \midrule    
    $\subscriber(\cp,\ci,\topic)$ & component $\cp$ subscribers to the topic $\topic$ through input port $\ci$.\\
     \midrule    
    $\informationflow(\ecu,\ci,\topic) $ & topic $\topic$ is published within $\ecu$ through an information flow from $\ci$. \\    
     \midrule
    $\protection(\topic)$ &  topic $\topic$ is protected by a cryptographic primitive.\\
     \midrule    
    $\public(\elem,\publico)$ & public $\elem$ and its output port $\publico$.\\
     \midrule    
    $\reach(\elem)$ & element $\elem$ is reachable by the intruder.\\ 
     \midrule    
    $\attack(\elem)$ & $\elem$ may be attacked by the intruder.\\     
    \bottomrule
  \end{tabular}
  \caption{Description of the predicates used to define the intruder's capabilities.}
  \label{tab:intruder_predicates}
\end{table}

We formalize an SOA intruder model defined by the rules in Figure~\ref{fig:intruder_capabilities}.
The intruder model is based on the main attacks against vehicle SOA with centralized architecture, described in Section~\ref{subsec:attacks-soa}. 
Intuitively, SOA contain two main attack surfaces that may be exploited if no suitable countermeasures are deployed.

\begin{itemize}
  \item \emph{Outsider Attackers} can exploit public interfaces, such as sensors and communication interfaces, to infiltrate the system and attack vehicle assets, such as safety functions.
  For example, attackers can spoof GPS coordinates thus violating the integrity of published position information by localization. 
  \item \emph{Insider Attackers} can exploit vulnerabilities in the underlying SOA protocols and carry out MITM attacks thus violating the integrity of topics.
  For example, attackers can carry out MITM attacks between localization and perception to violate the integrity of position information.
\end{itemize}

Figure~\ref{fig:intruder_capabilities} introduces the rules of the intruder model reflecting these type of attacks.
These inference rules derive three judgments described below. $\Gamma$ contains system specifications which are extracted from the vehicle model.
These specifications are formalized as atomic formulas using the predicate symbols described in Table~\ref{tab:intruder_predicates}.

$\Gamma \vdash \wrt(X,Y)$ and $\Gamma \vdash \rdg(X,Y)$ denote that the port $X$ of model element may write, respectively, read on $Y$. 
Rule $\mywrite_1$ specifies that an output $\eo$ of an ECU may write on an input port $\nti$ of a network element if an output port $\co$ of a component is allocated to $\eo$ (specified by $\cpo(\cp,\co), \alloc(\co,\eo)$), and there is a channel from $\eo$ to $\nti$ (specified by $\trans(\eo,\nti)$).
Rule $\mywrite_4$ is similar, but for public elements. 
Rule $\mywrite_2$ specifies that an input port of an ECU may write to its own output port -- we assume that there exists an internal transmission within the ECU ($\trans(\ei,\eo)$), \eg, components exchanging messages within the ECU.
Rule $\mywrite_3$ is similar, but for network interfaces.
Rule $\myread_2$ specifies when an ECU reads from a network interface (similar to $\mywrite_1$). 
Rule $\myread_1$ specifies that subscriber ports may read from publisher ports. 


$\Gamma \vdash \reach(X)$ denotes when a port $X$ of a model element is reachable by an intruder. 
  Rule $\basicpublic$ specifies that any port of a public element in the architecture can be reached by the (outsider) intruder.
  Rule $\inductivecaseonepublic$ specifies that a port $\porttwo$ of a model element can be reached by the (outsider) intruder if a port $\portone$ writes on $\porttwo$.
  Respectively, $\inductivecasetwopublic$ specifies that a port $\porttwo$ of a model element can be reached by the (outsider) intruder if $\porttwo$ reads on a port $\portone$.
  Rule $\basicsoa$ specifies that any publisher port in the architecture can be reached by the (insider) intruder.
  Rule $\inductivecaseonesoa$ specifies that the (insider) intruder can reach a subscriber port $\ci$ if $\ci$ reads on a reached publisher port $\co$. 

 $\Gamma \vdash \attack(X)$ denotes when a topic $X$ can be attacked. 
Rule $\attackoutside$ specifies that any topic published within an information flow ($\informationflow(\ecu,\port,\topic)$) from a reached ECU's input port may be attacked. 
  Rule $\attackinside$ specifies that any topic between publisher and subscriber ports reached by the (insider) intruder may be attacked if the topic is not protected.
\paragraph{Outsider Intruder (Example).} 

\begin{figure*}
\includegraphics[width=\textwidth]{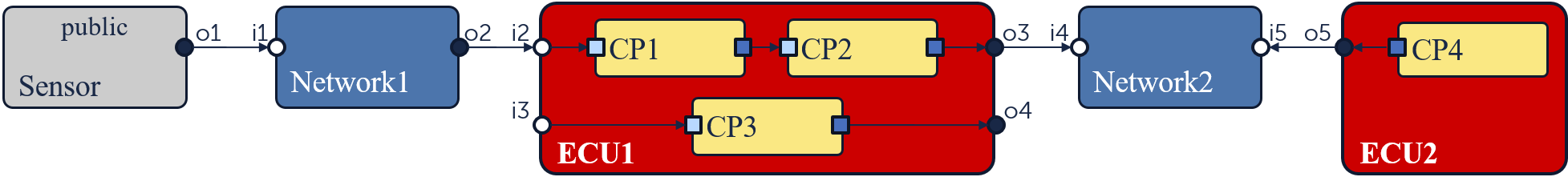}
\caption{Illustration of the outsider intruder}
\label{fig:intruder_model_public_interface}
\vspace*{0.3cm}
\includegraphics[width=\textwidth]{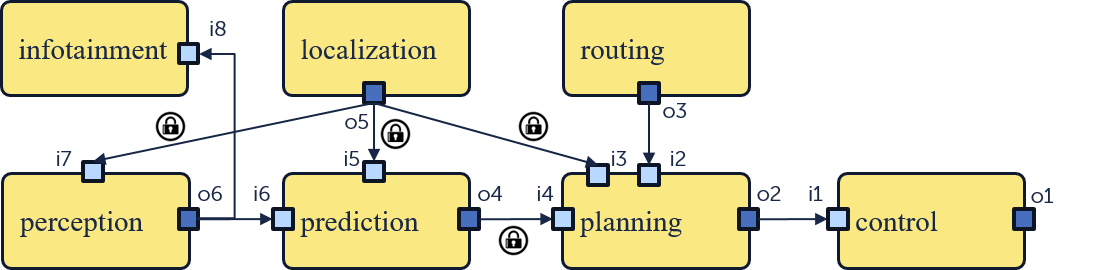}
\caption{Illustration of the insider intruder }
\label{fig:intruder_model_soa}
\end{figure*}

Consider the platform architecture depicted in Figure~\ref{fig:intruder_model_public_interface}.
The black and white circles connected to hardware units are, respectively, output and input ports.
We assume that \texttt{Sensor} is a public interface.
The output port \texttt{o1} of \texttt{Sensor} can be reached by the intruder based on the rule $\basicpublic$.
The output port \texttt{o1} writes on the input port \texttt{i1} of the network interface \texttt{Network1}, then based on $\inductivecaseonepublic$ the intruder can reach \texttt{i1} and \texttt{o2}.
We assume that the subscriber port (light blue square) of component \texttt{CP1} is allocated to the input port \texttt{i2} of \texttt{ECU1}, and that \texttt{i2} reads from \texttt{o2}.
The intruder can then reach \texttt{i2} and \texttt{o3} based on $\inductivecasetwopublic$.
Neither \texttt{i3} nor \texttt{o4} can be reached by the intruder.
The intruder can reach \texttt{i4} as \texttt{o3} writes to \texttt{i4}.
The intruder cannot reach \texttt{i5} and \texttt{o5}.
Finally, an intruder may carry out, \eg, a spoofing attack from \texttt{Sensor} to violate the integrity of the topics published by either \texttt{CP1} or \texttt{CP2} since there is an information flow from \texttt{i2} ($\attackoutside$).

\paragraph{Insider Intruder (Example).}
Consider the logical architecture depicted in Figure~\ref{fig:intruder_model_soa}.
The dark and light blue squares connected to components are, respectively, publisher and subscriber ports.
The intruder can reach all publisher ports \texttt{o1...o6} based on $\basicsoa$.
Based on $\inductivecaseonesoa$, the intruder can reach the subscriber ports \texttt{i1...i7}, as these ports read from publishers, \eg, \texttt{i7} reads from \texttt{localization} via port \texttt{o5}.
The intruder cannot reach the subscriber port \texttt{i8}, as \texttt{infotainment} is not a publisher.
We assume the topics published by ports \texttt{o4} and \texttt{o5} are protected.
Assume the topic published by \texttt{planning} through port \texttt{o2} is the intruder's target. 
As a result, the intruder has the following options to carry out MITM attacks.
An attack may be carried out between \texttt{routing} and \texttt{planning} or even between \texttt{perception} and \texttt{prediction} given that the topic published by \texttt{perception} may affect the topic published by \texttt{planning}.
The intruder can neither carry out attacks between \texttt{localization} and \texttt{planning} (same for \texttt{perception} and \texttt{prediction}), nor between \texttt{prediction} and \texttt{planning} since the topics are protected ($\attackinside$).
The intruder cannot carry out attacks from \texttt{infotainment}.

%% file: sec-automation.tex


\begin{table}[h]
\centering
\begin{tabular}{p{1.5cm}p{2cm}p{2.7cm}}
\toprule
 \multicolumn{1}{l}{Intruder}    & \#Attack Paths  & Execution time (s) \\ 
\midrule
 Outsider & 152 & 1.11 \\ 
 Insider & 94 & 0.06 \\ \bottomrule
 \end{tabular}
\caption{Number of identified attack paths and the execution time taken by \soamachinery to computed the attack paths.} 
\label{table:number_paths_execution_time}
\end{table}

\begin{table*}[t]
\centering
\begin{tabular}[t]{p{3.3cm}p{3.7cm}p{3.1cm}p{2.6cm}p{1cm} }
\toprule
\textbf{From} & \textbf{To} & \textbf{Affected Topic} & \textbf{Article} & \textbf{\#Attack Paths}\\
\toprule
\multicolumn{4}{c}{\textbf{Outsider Intruder}}\\
\midrule
\texttt{Bluetooth} & \texttt{VDC} & \texttt{signal} & \tiny{\cite{ChowdhuryKKJD20}} &  3 \\
\texttt{LiDAR} & \texttt{MCU} & \texttt{obstacles} & \tiny{\cite{abs-2102-03722}} &  24  \\
\texttt{Front Left Camera} & \texttt{MCU} & \texttt{obstacles} & \tiny{\cite{JhaCBCTKI20}} &  18 \\
\texttt{GPS} & \texttt{MCU} & \texttt{localization pose} & \tiny{\cite{Shen20}} &  18 \\
\texttt{Front Radar} & \texttt{MCU} & \texttt{obstacles} & \tiny{\cite{KomissarovW21}} &  6  \\
\texttt{T-Box} & \texttt{MCU} & \texttt{traffic light} & NA &  18  \\
\toprule
\multicolumn{4}{c}{\textbf{Insider Intruder}}\\
\midrule
\texttt{gnss driver} & \texttt{velodyne detection} & \texttt{tf} & \tiny{\cite{HongKJCCMM20}} & 1  \\
\texttt{gnss driver} & \texttt{msf localization} & \texttt{gnss best pose}  & \tiny{\cite{HongKJCCMM20}} & 1 \\
\texttt{compensator} & \texttt{velodyne detection} & \texttt{pointcloud2}  & \tiny{\cite{HongKJCCMM20}} & 1 \\
\texttt{control} & \texttt{chassis} & \texttt{signal} & \tiny{\cite{HongKJCCMM20}} & 1 \\
\texttt{chassis} & \texttt{gnss driver} & \texttt{chassis} & \tiny{\cite{HongKJCCMM20}} & 1 \\
\texttt{v2x proxy} & \texttt{traffic light} & \texttt{traffic light} & NA & 1 \\
\texttt{routing} & \texttt{planning} & \texttt{routing response} & NA & 1 \\
\texttt{relative map} & \texttt{planning} & \texttt{map} & NA & 1 \\
\bottomrule
\end{tabular}
\caption{Potential attacks derived from selected attack paths. \textit{From} and \textit{To} denote, respectively, the model element where the attack starts and ends. \textit{Affected Topic} denotes the actual target of the attacker. The upper and lower part of the table describe, respectively, selected attacks carried out by the outsider and insider intruder, incl. attacks reported in the literature. NA denotes attacks that up to the best of our knowledge have not been reported in the literature. \textit{\#Attack Paths} denotes that number of computed attack paths \textit{From} and \textit{To}.}
\label{table:insider_intruder_attack_paths}
\end{table*}

\section{Automating Attack Path Analysis}\label{sec:automation}


\soamachinery (vehic\textbf{L}e thre\textbf{A}t analysis a\textbf{U}tomation \textbf{F}or s\textbf{E}rvice-orie\textbf{N}ted architectures) is an SOA machinery that enables the automated computation of several activities of the 
Threat Assessment and Remediation Analysis (TARA) analysis.
Based on our safe and secure-by-design methodology, \soamachinery can compute assets, damage scenarios, impact rating, threat scenarios, and attack paths.
This section focuses on the automated computation of attack paths that can cause threat scenarios to vehicle SOA, \ie, the paths that violate cybersecurity properties of assets (Section~\ref{sec:safety_analysis}).
To this end, \soamachinery implements the proposed intruder model in the logic programming tool DLV~\cite{leone06tcl}.
Besides being declarative and expressive enough to implement the intruder model for vehicle SOA, 
logic programming methods are well-known to be suitable for reasoning about paths, such as path reachability~\cite{baral.book}.
\soamachinery encodes the system specification as facts using the predicates described in Table~\ref{tab:intruder_predicates}, and the intruder model described in Section~\ref{sec:intruder-model}.
Then the DLV solver is used to enumerate the attack paths.
We validate \soamachinery on the modeled Apollo system architecture. 
The implementation and the experimental results are available at~\cite{laufen_machinery}.

Given the high complexity of the Apollo model, naively computing the attack paths based on reachability does not scale, in particular for the outsider intruder.
To address this problem, the computation is divided into two steps.
The first step, \emph{Intruder reachability}, computes all the model elements that are reachable by the intruder as specified by the write and read, and reachability rules.
Since no paths are computed, the DLV engine computes the reachable elements in the range of milliseconds. 
We then use the reachable elements as input to the second step, \emph{Path computation}, where we 
make use of the attack rules. 
Instead of enumerating all paths, we proceed using a goal-oriented search to enumerate only the attack paths on assets (a.k.a. asset-centric approach).
This means that DLV does not require to compute all paths.

We run the experiments on a 1.90GHz Intel Core i7-8665U with 16GB of RAM running Ubuntu 18.04 LTS with kernel 5.4.0-113-generic and DLV 2.1.1.
Table~\ref{table:number_paths_execution_time} shows the number of identified attack paths, and the execution time of \soamachinery.
The execution time in enumerating the attack paths is rather low, \ie, 1.11 and 0.06 seconds for the outsider and insider intruder, respectively.  
The number of identified attack paths is high due the complexity of the system, \eg, the great number of public elements and the great number of information flows in the architecture.
We do not rule out any attack path to guarantee a complete coverage of possible steps exploited by the intruder. 
Section~\ref{sec:security_countermeasures} elaborates on countermeasures that may mitigate several of the identified attack paths.

We analyzed the generated attack paths w.r.t. potential attacks against safety-critical topics.
Table~\ref{table:insider_intruder_attack_paths} organizes selected attack paths into attacks carried out by an outsider attacker and insider attacker.


Firstly, our analysis was able to identify several attacks that have been reported in the literature, namely those attacks associated with a citation. 
The set of attack paths computed includes attacks carried out by outsider attackers exploiting \texttt{Bluetooth}, \texttt{LiDAR}, \texttt{Camera}, \texttt{GPS} and \texttt{Radar} that may target safety-critical topics, cause loss scenarios and harm to road-users, as well as by insider attackers exploiting SOA communication vulnerabilities to target topics.

Secondly, we also identified potential attacks that up to the best of our knowledge have not been reported in the literature (marked with NA). 
We analyzed in further detail some of the attacks by using the model and its connection to the Apollo code to find out how such attacks can lead to safety problems.

\begin{figure}[H]
\begin{lstlisting}[style=CStyle,basicstyle=\tiny]
void TrafficLightsPerceptionComponent::OnReceiveImage(
    const std::shared_ptr<apollo::drivers::Image> msg,
    const std::string& camera_name) { ...
 /** Set traffic light status based on camera data **/
  traffic_light_pipeline_->Perception(
  camera_perception_options_,frame_.get()); ...
 /** Overwrites traffic light status if valid v2x data **/
  SyncV2XTrafficLights(frame_.get()); ...}
\end{lstlisting}
\vspace*{-0.2cm}
\caption{Code snippet of Apollo: Overwriting traffic light status with V2X data.}
\label{fig:code_snippet_traffic_light_component}
\end{figure}

\paragraph{V2X Traffic Light Overwrite Attack.}
\begin{figure*}[t]
\begin{center}
\includegraphics[width=\textwidth]{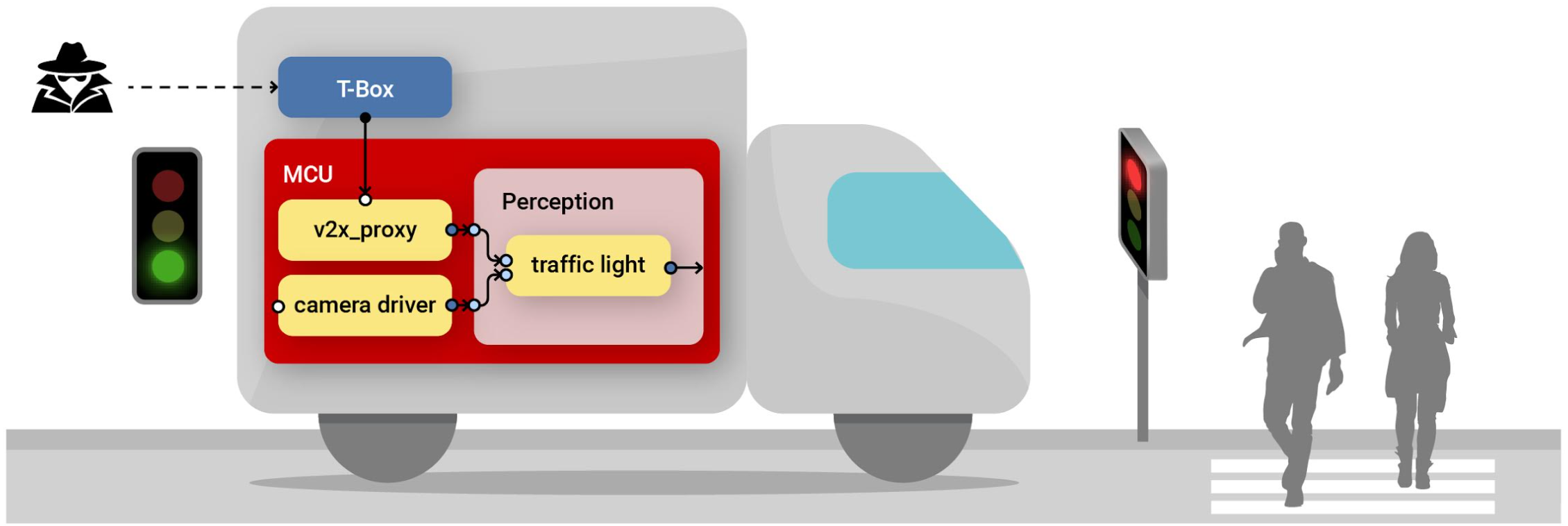}  
\end{center}
\caption{Illustration of a spoofing attack from \texttt{T-Box} to manipulate traffic light status received by \texttt{v2v proxy}.} 
\label{fig:t_box_traffic_light_attack}
\end{figure*}
\soamachinery has identified attack paths targeting \texttt{v2v proxy} from both outside and inside.
Figure~\ref{fig:t_box_traffic_light_attack} illustrates the attack from the outside.
The \texttt{v2v proxy} component publishes data on the traffic light status obtained from the road infrastructure. 
This data is subscribed by the \texttt{traffic light} component which also subscribes data from the cameras to identify and publish the traffic light status.
Since there is a traceability from the Apollo model and the Apollo source code, it is straightforward to find the relevant classes for vulnerabilities.
Indeed, we found out that the function TrafficLightsPerceptionComponent gives priority to the data received by \texttt{v2v proxy} over the data received by the cameras.
Figure~\ref{fig:code_snippet_traffic_light_component} shows a code snippet from the TrafficLightsPerceptionComponent function.
As a result, an attacker may manipulate the traffic light status from either outside (\ie, spoofing attack) or inside (\ie, MITM attack). 
As illustrated by Figure~\ref{fig:t_box_traffic_light_attack}, a spoofing attack from \texttt{T-Box} manipulating the traffic light status can cause serious harm to passengers and pedestrians as the vehicle can cross a red-light. 




\paragraph{Route/Mapping Injection Attack.}
The seriousness of some of the identified attack paths may not be too obvious from a safety perspective.
For example, \soamachinery has identified the following attack path: \texttt{\{routing, planning\}} targeting the \texttt{routing response} topic.
An insider attacker may carry out a MITM attack between \texttt{routing} and \texttt{planning} to provide a malicious route for the ego vehicle.
From a safety perspective, a loss scenario of type erroneous from \texttt{routing} to \texttt{planning} may be easy to control, and hence it would lead to a low criticality hazard (\eg, ASIL A or B).
From a security perspective, however, there are several serious consequences, including hijacking of passengers.
The \texttt{planning} component may also be affected by the road map published by \texttt{relative map}, as \texttt{planning} takes the map into account while computing the vehicle trajectory.

\subsection{Potential Countermeasures}\label{sec:security_countermeasures}
We have performed an attack path analysis to deduce potential locations for instantiating security countermeasures.
Our analysis focused on the computed attack paths using the outsider intruder.
This choice was made because we noticed that many of such attack paths have the same prefix, which may be a hint for instantiating security countermeasures.

Table~\ref{table:attack_path_analysis} presents the main results of our attack path analysis.
Specifically, Table~\ref{table:attack_path_analysis} shows the public element that can be reached the outsider intruder, the number of attack paths computed by \soamachinery from the public element, and the common prefix for all attack paths from the same public element.

\begin{table*}[h]
\centering
\begin{tabular}{p{5cm}p{3cm}p{6cm}}
\toprule
 \multicolumn{1}{l}{\textbf{Public element}}    & \textbf{\#Attack Paths}  & \textbf{Prefix} \\ 
\midrule
 \texttt{Front Left Camera} & 21 & \texttt{Front Left Camera} $\rightarrow$ \texttt{GMSL} $\rightarrow$ \texttt{VIU 1} \\ 
\midrule
 \texttt{Front Right Camera} & 21 & \texttt{Front Right Camera} $\rightarrow$ \texttt{GMSL} $\rightarrow$ \texttt{VIU 3} \\ 
\midrule
 \texttt{GPS} & 21 & \texttt{GPS} $\rightarrow$ \texttt{Serial} $\rightarrow$ \texttt{VIU 3} \\ 
\midrule
 \texttt{Front Radar} & 10 & \texttt{Front Radar} $\rightarrow$ \texttt{CAN} $\rightarrow$ \texttt{MDC}\\ 
\midrule
 \texttt{Rear Radar} & 10 & \texttt{Rear Radar} $\rightarrow$ \texttt{CAN} $\rightarrow$ \texttt{MDC}\\  
\midrule
 \texttt{LiDAR} & 27 & \texttt{LiDAR} $\rightarrow$ \texttt{SW4} \\   
 \midrule 
 \texttt{Bluetooth} & 21 & \texttt{Bluetooth} $\rightarrow$ \texttt{USB} $\rightarrow$ \texttt{CDC} \\
 \midrule 
 \texttt{T-Box} & 21 & \texttt{T-Box} $\rightarrow$ \texttt{SW3}  \\ 
\bottomrule
 \end{tabular}
\caption{Attack paths analysis (outsider intruder)}
\label{table:attack_path_analysis}
\end{table*}

The last architecture element described in the Prefix column may be a suitable location to instantiate a countermeasure and consequently address the attack paths. 
From the Prefix column, we can also notice that the gateway \texttt{VIU 3} is a common location in the attack paths from both \texttt{Front Right Camera} and \texttt{GPS}.
Similarly, the connection between \texttt{CAN} and \texttt{MDC} is a common location in the attack paths from \texttt{Front Radar} and \texttt{Rear Radar}.
This gives us a hint that security countermeasures could be placed in front of \texttt{VIU 3}, and between \texttt{CAN} and \texttt{MDC} to address such attack paths (specifically, 62 attack paths).

Firewalls are, \eg, recommended~\cite{ChengDPP19} as means to protect vehicle architectures against such attacks. They may be deployed in front of the last architecture elements in the Prefix column to filter network traffic and prevent malicious intrusion.
For the network interfaces (\ie, \texttt{T-Box} and \texttt{Bluetooth}), one could also implement a mutual authentication mechanism (\eg, mTLS) to ensure that only authenticated messages are accepted.

Safety architecture patterns, such as Heterogeneous Duplex pattern~\cite{Armoush2010}, may also be deployed as a second-layer of defense. 
Consider, \eg, the V2X traffic light overwrite attack carried by an outsider attacker.
This attack violates the integrity of \texttt{traffic light} topic through \texttt{T-Box}.
A possible countermeasure is to include a checker in the \texttt{traffic light} component to consider inputs from both \texttt{v2x proxy} and cameras (\ie, heterogeneous inputs) -- the \texttt{traffic light} component emits an alert to the driver or transition the system to a safe state if the inputs do not match.

The MITM attacks (\eg, the route injection attack) carried by an insider attacker exploit SOA communication vulnerabilities to violate the integrity of topics.
Digital signatures are a well-known countermeasure for ensuring authenticity and integrity between servers (\eg, publishers) and clients (\eg, subscribers).
To address MITM attacks, one can implement digital signatures in the Apollo system, where each publisher originator signs its message, and each subscribe of the message verifies the signature of the message.
Fast-DDS provides a cryptographic plugin for message authentication codes computation and verification.
The use of digital certificates to address MITM attacks in Apollo was inspired by \cite{HongKJCCMM20} that proposed a countermeasure using digital signatures for mitigating publisher-subscriber overprivilege issues in Apollo.

All 94 attack paths (insider intruder) are, in principle, addressed upon implementing digital signatures.
The decision of using digital signatures causes, however, a performance penalty at the execution time of software components.
The performance penalty can then be analyzed according to several points of view, including security, safety and financial.
However, since all of the identified attack paths are safety critical, countermeasures shall be implemented to ensure vehicle safety.

%% file: sec-related_work.tex
\section{Related Work}\label{sec:related_work}


\paragraph{Safe and Security by Design.}
There is a rich literature in safety and security co-design on which our methodology is built upon. 
We detail some key approaches that are more closely related comparing them with our approach.

System-theoretic Process Analysis for Security (STPA-SEC)~\cite{YoungL13} is an extension of the STPA method to compute both safety artifacts and security artifacts (\ie, vulnerabilities).
STPA and Six Step Model~\cite{Sabaliauskaite18} is an approach that integrates safety and security artifacts for autonomous vehicles.
The approach uses STPA and the Six-Step Model to specify safety and security artifacts, in particular threats are derived from failures that lead to hazardous events identified in a Hazard Analysis and Risk Assessment (HARA) analysis.
The Safety-Aware Hazard Analysis and Risk Assessment (SAHARA)~\cite{MacherSBAK15} approach extends the HARA analysis of ISO 26262~\cite{iso26262} to include security threats that may have a safety impact.
The security threats are derived by SAHARA with the help of STRIDE.
The STRIDE methodology~\cite{shostack14book} is a well-known threat modeling proposed by Microsoft.  
STRIDE represents six types of threats, namely \textbf{S}poofing, \textbf{T}ampering, \textbf{R}epudiation, \textbf{I}nformation disclosure, \textbf{D}enial of service, and \textbf{E}levation of privilege.
These type of threats can be derived from the security property that the system shall satisfy, \eg, tampering can be derived from the integrity property.
The Bosch engineers~\cite{bosch-paper} proposed an approach for deriving security artifacts of Threat Analysis and Risk Assessment (TARA) from safety artifacts computed by a HARA analysis.
Specifically, the Bosch approach recommends that 
(1) assets are derived from safety goals, 
(2) threats are derived from the violation of safety goals, 
(3) damage scenarios are derived from hazards, and 
(4) impact rating values are derived from severity/controllability of ASIL.

Our safe and secure-by-design method illustrated in Figure~\ref{fig:approach} is inspired by the above methods/approaches.
Following the Bosch approach, our method expects a safety analysis to be first performed.
Then based on the results of the safety analysis, our method derives security artifacts.
We also agree with the Bosch approach regarding deriving damage scenarios from hazards, and impact rating values from severity/controllability of ASIL.
Our method considers functions, topics, and hardware units as assets that shall also be protected from a security perspective.
We do not see how such assets can be derived from safety goals as recommended by the Bosch approach.
Therefore, similar to STPA-SEC and STPA and Six Step Model, our safe and security-by-design method considers the results of STPA (in addition to HARA).
That is, our method derives assets from the loss scenarios computed by STPA.
Similar to SAHARA and also recommended by ISO 21434~\cite{whitepaper-sembera}, we model threats based on the STRIDE methodology.
That is, we derive (a) the security property that the system (\eg, a function) shall satisfy from the failure mode associated to a loss scenario, and (b) the threat type from the desired security property, in particular our work focuses on spoofing and tampering threats that may violate the integrity of safety-critical topics.

\paragraph{Attacks Against Vehicle SOA.}
\label{subsec:attacks-soa}

The following work has inspired us to formalize the intruder model for vehicle SOA.
A recent systematization of knowledge article~\cite{Shen22} gives an overview of the state-of-the-art of the literature.
The article analyzed 53 articles and taxonomize them based on security critical aspects, including attacks against sensors.
In ``Drift with the devil'' \cite{Shen20} it is shown that an intruder may manipulate location information by spoofing GPS radio signals.
This attack is effective even against localization components using multi-sensor fusion.
LiDAR sensor signals may be spoofed to remove obstacles on the road~\cite{abs-2102-03722}.
Camera signals may also be spoofed to manipulate video frames given that the camera traffic is transmitted in plain text~\cite{JhaCBCTKI20}.
Attackers may carry out spoofing attacks to inject signals into a radar sensor to make it perceive fake obstacles~\cite{KomissarovW21}.
An attack may exploit vulnerabilities in a Bluetooth stack weakness to lock the brakes of the vehicle~\cite{ChowdhuryKKJD20}.
In the work by~\cite{ZelleLKK21}, the authors investigate possible security issues in the service discovery mechanism of vehicle SOA, in particular SOA using the SOME/IP protocol, enabling the attacker to carry out MITM attacks between publishers and subscribers.
In the work on AVGuardian~\cite{HongKJCCMM20}, the authors investigated possible publisher/subscriber overprivilege instances in Apollo.
The AVGuardian tool detected several overprivilege instances in the Apollo 5.0 code base, including overprivilege instances in (a) the gnss driver that may exploit a publish-overprivileged field in the tf topic to relocate the estimated position of a perceived obstacle in the road and (b) the compensator that may exploit a publish-overprivileged filed in the PointCloud topic to remove a perceived obstacle from the road. 

Our intruder model specifies the main attacker's capabilities needed to carry out the above attacks at the architecture level, including the capabilities of attackers to carry out (a) spoofing attacks from outside (\eg, from sensors), and (b) MITM attacks from inside (\eg, between components), thus violating the integrity of safety-critical topics.
The attacks exploiting overprivileged instances can be seen as a specific case of the MITM attack.

\paragraph{Automate Threat Analysis.}
To date, not many tools provide computed-aided support for computing threats and attack paths. 
A survey on threat modeling~\cite{XiongL19} has shown that most threat modeling work remains to be done manually.
We briefly describe some of the security/threat analysis tools that provide computed-aided support in the automotive domain.

AVGuardian~\cite{HongKJCCMM20} is a static analysis tool to detect overprivilege instances in source code implementing service-oriented architectures for automotive systems.
AVGuardian examines each module's source code and automatically detects publisher and subscriber overprivilege instances in the fields of topics defined by the module.
AVGuardian requires the behavior specification of the system to detect overprivilege instances.
\soamachinery has been implemented to identify threats and attack paths during the design of the system architecture without the behavior specification of the system.
\soamachinery has been able to automatically compute attack paths that may lead to the attacks detected by the AVGuardian tool. 
We agree that with the behavior specification one can obtain more accurate information w.r.t. assets and potential attacks, \eg, which field of the topic is relevant for the overprivilege instance. 

ProVerif~\cite{pro_verif} and Tamarin Prover~\cite{tamarin_prover} are well-known automated reasoning tools to verify the security properties of systems (in particular, security protocols) with the Dolev-Yao intruder model~\cite{DY}.
These reasoning tools require the formal specification of the behavior of the system to verify its properties.
A promising future work direction is to include the behavior specification in our Apollo model and use such reasoning tools to verify security properties of SOA protocols such as SOME/IP or DDS.

Previous works~\cite{nigam22jlamp,apvrille15ccis} propose formal threat analysis using models of cyber-physical systems, such as for Industry 4.0 applications. 
Similar to the work on security protocols, these works require the formal specification of the behavior of the system. As investigated in~\cite{nigam22jlamp}, these methods have scalability limitations due to the state-space problem, as the time of analysis increases exponentially with the number of components.
It is, therefore, unlikely that such methods alone will scale to the size of the Apollo system with more than 60 components. We believe that an interesting future work is to combine our threat analysis methods that identifies attack paths with methods that reason using the formal specification of the behavior, so to provide the precision of the analysis of methods that use the formal behavior with the scalability of our methods.

An attack propagation method that targets automotive safety-critical functions has been proposed by~\cite{FockelSTSK22}.
The commercial tool ThreatGet~\cite{threatget} enables the identification of attack paths following the ISO 21434 standard.
Microsoft SDL Threat Modeling tool~\cite{sdl_threat_modeling_tool} is another well-known commercial tool to compute threats.
The threats are computed using the STRIDE methodology.
The attack path associated to each compute threat is represented using data flow diagrams.
To the best of our knowledge, these tools do not support intruder model capabilities for vehicle SOA.
As a result, we advance the state-of-the-art by proposing a machinery built upon realistic formalized intruder models for vehicle SOA.

%% file: sec-conclusion.tex
\vspace*{-7mm}
\section{Conclusion}\label{sec:conclusion}
\vspace*{-3mm}

This article proposed automated methods for threat analysis using a model-based engineering methodology.
To this end, we have (a) modeled a faithful vehicle model of the autonomous driving functions of the Apollo framework, (b) formalized an intruder model for vehicle SOA based on the literature review, and (c) developed \soamachinery, an SOA machinery for computing several activities of a threat analysis, including attack paths.

%% file: icissp23.bbl
\begin{thebibliography}{}

\bibitem[sae, 2018]{sae}
 (2018).
\newblock {SAE International Releases Updated Visual Chart for Its ``Levels of
  Driving Automation'' Standard for Self-Driving Vehicles}.

\bibitem[cyb, 2019]{cyberrt}
 (2019).
\newblock {Apollo Cyber RT}.
\newblock Available at \url{https://cyber-rt.readthedocs.io/}.

\bibitem[sdl, 2022]{sdl_threat_modeling_tool}
 (2022).
\newblock {Microsoft SDL Threat Modeling Tool}.
\newblock Available at
  \url{https://www.microsoft.com/en-us/securityengineering/sdl/threatmodeling}.

\bibitem[thr, 2022]{threatget}
 (2022).
\newblock {{ThreatGet} - Threat Analysis and Risk Management}.
\newblock Available at \url{https://www.threatget.com/}.

\bibitem[Apollo, 2021]{apollo}
Apollo (2021).
\newblock {An Open Autonomous Driving Platform}.
  \url{https://github.com/ApolloAuto/apollo}.

\bibitem[Apvrille and Roudier, 2015]{apvrille15ccis}
Apvrille, L. and Roudier, Y. (2015).
\newblock {Designing Safe and Secure Embedded and Cyber-Physical Systems with
  SysML-Sec}.
\newblock In {\em {MODELSWARD'15}}.

\bibitem[Aravantinos et~al., 2015]{Aravantinos2015}
Aravantinos, V., Voss, S., Teufl, S., H{\"o}lzl, F., and Sch{\"a}tz, B. (2015).
\newblock {{AutoFOCUS~3}: Tooling Concepts for Seamless, Model-based
  Development of Embedded Systems}.
\newblock In {\em {ACES-MB'15}}.

\bibitem[Armoush, 2010]{Armoush2010}
Armoush, A. (2010).
\newblock {\em Design Patterns for Safety-Critical Embedded Systems}.
\newblock PhD thesis, {RWTH} Aachen University.

\bibitem[Baral, 2010]{baral.book}
Baral, C. (2010).
\newblock {\em Knowledge Representation, Reasoning and Declarative Problem
  Solving}.
\newblock Cambridge University Press.

\bibitem[Basin et~al., 2022]{tamarin_prover}
Basin, D., Cremers, C., Dreier, J., Meier, S., Sasse, R., and Schmidte, B.
  (2022).
\newblock {Tamarin Prover} \url{https://tamarin-prover.github.io/}.

\bibitem[Blanchet et~al., 2022]{pro_verif}
Blanchet, B., Cheval, V., Allamigeon, X., Smyth, B., and Sylvestre, M. (2022).
\newblock {ProVerif} \url{https://bblanche.gitlabpages.inria.fr/proverif/}.

\bibitem[Cheng et~al., 2019]{ChengDPP19}
Cheng, B. H.~C., Doherty, B., Polanco, N., and Pasco, M. (2019).
\newblock {Security Patterns for Automotive Systems}.
\newblock In {\em {MODELS'19}}.

\bibitem[Chowdhury et~al., 2020]{ChowdhuryKKJD20}
Chowdhury, A., Karmakar, G.~C., Kamruzzaman, J., Jolfaei, A., and Das, R.
  (2020).
\newblock {Attacks on Self-Driving Cars and Their Countermeasures: {A} Survey}.
\newblock {\em {IEEE} Access}.

\bibitem[Dantas and Nigam, 2022a]{dantas_journal_safety_security_patterns}
Dantas, Y.~G. and Nigam, V. (2022a).
\newblock {Automating Safety and Security Co-Design through Semantically-Rich
  Architectural Patterns}.
\newblock {\em {ACM} Trans. Cyber Phys. Syst.}

\bibitem[Dantas and Nigam, 2022b]{laufen_machinery}
Dantas, Y.~G. and Nigam, V. (2022b).
\newblock \url{https://github.com/ygdantas/LAUFEN}.

\bibitem[Dolev and Yao, 1983]{DY}
Dolev, D. and Yao, A.~C. (1983).
\newblock On the security of public key protocols.
\newblock {\em {IEEE} Trans. Inf. Theory}, 29(2):198--207.

\bibitem[Fockel et~al., 2022]{FockelSTSK22}
Fockel, M., Schubert, D., Trentinaglia, R., Schulz, H., and Kirmair, W. (2022).
\newblock Semi-automatic integrated safety and security analysis for automotive
  systems.
\newblock In {\em {MODELSWARD'22}}.

\bibitem[F{\"{o}}rster et~al., 2019]{bosch-paper}
F{\"{o}}rster, D., Loderhose, C., Bruckschl{\"{o}}gl, T., and Wiemer, F.
  (2019).
\newblock Safety goals in vehicle security analyses: a method to assess
  malicious attacks with safety impact.
\newblock In {\em the 17th escar Europe - Embedded Security in Cars}.

\bibitem[{fortiss GmbH}, 2022]{af3}
{fortiss GmbH} (2022).
\newblock {AutoFOCUS 2.21}.

\bibitem[Hau et~al., 2021]{abs-2102-03722}
Hau, Z., Co, K.~T., Demetriou, S., and Lupu, E.~C. (2021).
\newblock Object removal attacks on lidar-based 3d object detectors.
\newblock {\em CoRR}, abs/2102.03722.

\bibitem[Hong et~al., 2020]{HongKJCCMM20}
Hong, D.~K., Kloosterman, J., Jin, Y., Cao, Y., Chen, Q.~A., Mahlke, S.~A., and
  Mao, Z.~M. (2020).
\newblock {AVGuardian: Detecting and Mitigating Publish-Subscribe Overprivilege
  for Autonomous Vehicle Systems}.
\newblock In {\em {EuroS{\&}P'20}}.

\bibitem[ISO26262, 2018]{iso26262}
ISO26262 (2018).
\newblock {ISO 26262}, road vehicles — functional safety — part 6: Product
  development: software level.

\bibitem[{{ISO/SAE} 21434}, 2020]{iso21434}
{{ISO/SAE} 21434} (2020).
\newblock Road vehicles - cybersecurity engineering.

\bibitem[Jha et~al., 2020]{JhaCBCTKI20}
Jha, S., Cui, S., Banerjee, S.~S., Cyriac, J., Tsai, T., Kalbarczyk, Z., and
  Iyer, R.~K. (2020).
\newblock {ML-Driven Malware that Targets {AV} Safety}.
\newblock In {\em {DSN} 2020}.

\bibitem[Komissarov and Wool, 2021]{KomissarovW21}
Komissarov, R. and Wool, A. (2021).
\newblock Spoofing attacks against vehicular {FMCW} radar.
\newblock In {\em {ASHES@CCS'21}}.

\bibitem[Leone et~al., 2006]{leone06tcl}
Leone, N., Pfeifer, G., Faber, W., Eiter, T., Gottlob, G., Perri, S., and
  Scarcello, F. (2006).
\newblock The {DLV} system for knowledge representation and reasoning.
\newblock {\em {ACM} Trans. Comput. Log.}, 7.

\bibitem[Leveson and Thomas, 2018]{stpa-handbook}
Leveson, N.~G. and Thomas, J.~P. (2018).
\newblock {\em {STPA Handbook}}.

\bibitem[Macher et~al., 2015]{MacherSBAK15}
Macher, G., Sporer, H., Berlach, R., Armengaud, E., and Kreiner, C. (2015).
\newblock {{SAHARA:} A Security-aware Hazard and Risk Analysis Method}.
\newblock In {\em {DATE'15}}.

\bibitem[Nigam and Talcott, 2022]{nigam22jlamp}
Nigam, V. and Talcott, C.~L. (2022).
\newblock Automated construction of security integrity wrappers for industry
  4.0 applications.
\newblock {\em J. Log. Algebraic Methods Program.}

\bibitem[Sabaliauskaite et~al., ]{Sabaliauskaite18}
Sabaliauskaite, G., Liew, L.~S., and Cui, J.~C.

\bibitem[Sembera, 2020]{whitepaper-sembera}
Sembera, V. (2020).
\newblock {ISO/SAE} 21434: Setting the standard for connected cars’
  cybersecurity.
\newblock White Paper.

\bibitem[Shen et~al., 2022]{Shen22}
Shen, J., Wang, N., Wan, Z., Luo, Y., Sato, T., Hu, Z., Zhang, X., Guo, S.,
  Zhong, Z., Li, K., Zhao, Z., Qiao, C., and Chen, Q.~A. (2022).
\newblock Sok: On the semantic {AI} security in autonomous driving.
\newblock {\em CoRR}, abs/2203.05314.

\bibitem[Shen et~al., 2020]{Shen20}
Shen, J., Won, J.~Y., Chen, Z., and Chen, Q.~A. (2020).
\newblock {Drift with Devil: Security of Multi-Sensor Fusion based Localization
  in High-Level Autonomous Driving under {GPS} Spoofing}.
\newblock In {\em {USENIX'20}}.

\bibitem[Shostack, 2014]{shostack14book}
Shostack, A. (2014).
\newblock {\em Threat Modeling: Designing for Security}.
\newblock Wiley.

\bibitem[SOTIF, 2021]{sotif}
SOTIF, I.~. (2021).
\newblock {Safety of the Intended Functionality}.

\bibitem[UN, 2021]{unece}
UN (2021).
\newblock {UN Regulation No. 155 - Cyber security and cyber security management
  system}.

\bibitem[{WIRED}, 2015]{attack.jeep}
{WIRED} (2015).
\newblock Hackers remotely kill a jeep on the highway-with me in it.
\newblock Available at
  \url{https://www.wired.com/2015/07/hackers-remotely-kill-jeep-highway/}.

\bibitem[Xiong and Lagerstr{\"{o}}m, 2019]{XiongL19}
Xiong, W. and Lagerstr{\"{o}}m, R. (2019).
\newblock Threat modeling - {A} systematic literature review.
\newblock {\em Comput. Secur.}, 84:53--69.

\bibitem[Young and Leveson, 2013]{YoungL13}
Young, W. and Leveson, N.~G. (2013).
\newblock Systems thinking for safety and security.
\newblock In {\em {ACSAC'13}}.

\bibitem[Zelle et~al., 2021]{ZelleLKK21}
Zelle, D., Lauser, T., Kern, D., and Krau{\ss}, C. (2021).
\newblock {Analyzing and Securing {SOME/IP} Automotive Services with Formal and
  Practical Methods}.
\newblock In {\em {ARES'21}}.

\end{thebibliography}
